\documentstyle[12pt]{article}
\def\be{\begin{equation}}
\def\ee{\end{equation}}
\def\bea{\begin{eqnarray}}
\def\eea{\end{eqnarray}}
\def\bq{\begin{quote}}
\def\eq{\end{quote}}
\def\bseq{\begin{subequation}}

\def\eseq{\end{subequation}}
\def\bsea{\begin{subeqnarray}}

\def\esea{\end{subeqnarray}}
\def\sb{\sin\beta}
\def\cb{\cos\beta}
\def\cbb{\cos 2 \beta}

\def\deltarho{\Delta \rho (t,b) + \Delta \rho (\tilde{t},\tilde{b})}
\def\simlt{\stackrel{<}{{}_\sim}}
\def\simgt{\stackrel{>}{{}_\sim}}
\def\ov{\overline}

\def\tb{\tan\beta}

\def\ma{m_A}

\def\msqtb{(\tilde{m},\tan \beta)}

\def\msq{\tilde{m}}

\def\msba{m_{\tilde{b}_1}}

\def\msbl{m_{\tilde{b}_L}}
\def\mstl{m_{\tilde{t}_L}}
\def\st{\tilde{t}}

\def\gev{{\rm \; GeV}}

\begin{document}
\begin{titlepage}
\vspace*{-1cm}
\noindent
\phantom{DRAFT 27/3/93}
\hfill{hep-ph/9303317}
\\
\phantom{bla}
\hfill{CERN-TH.6833/93}
\\
\phantom{bla}
\hfill{IEM-FT-69/93}
\vskip 2.0cm
\begin{center}
{\Large\bf On the electroweak phase transition in the}
\end{center}
\begin{center}
{\Large\bf Minimal Supersymmetric Standard Model}
\end{center}
\vskip 1.5cm
\begin{center}
{\large J.R. Espinosa} \footnote{Supported by a grant of Comunidad
de Madrid, Spain.},
{\large M. Quir\'os} \footnote{Work partly supported by CICYT, Spain,
under contract AEN90-0139.}
\\
\vskip .3cm
Instituto de Estructura de la Materia, CSIC\\
Serrano 123, E-28006 Madrid, Spain\\
\vskip .3cm
and \\
\vskip .3cm
{\large F. Zwirner} \footnote{On leave from INFN, Sezione di Padova,
Padua, Italy.}
\\
\vskip .3cm
Theory Division, CERN \\
CH-1211 Geneva 23, Switzerland \\
\end{center}
\vskip 1cm
\begin{abstract}
\noindent
We study the finite-temperature effective potential of the Minimal
Supersymmetric Standard Model, in the limit of only one light Higgs
boson.
Because of the large top Yukawa coupling, there can be significant
differences with respect to the Standard Model case: for given values
of the Higgs and top masses, little supersymmetry breaking in the
stop sector can
make the phase transition more strongly first-order. After including
the full structure of the stop mass matrix, the most important
experimental constraints and the leading plasma effects, we find that
the present limits on Higgs and squark masses are still compatible
with the scenario of electroweak baryogenesis, in a small region of
parameter space corresponding to $m_h \simlt 70 \gev$ and $\msba
\simlt 105 \gev$.

\end{abstract}
\vfill{
CERN-TH.6833/93
\newline
\noindent
March 1993}
\end{titlepage}
\setcounter{footnote}{0}
\vskip2truecm

{\bf 1.}
The possibility of generating the cosmological matter-antimatter
asymmetry at the electroweak phase transition has been the subject of
lively discussions in the recent literature (for updated reviews and
bibliography, see ref.~[\ref{reviews}]). Many difficult problems must
be faced by any attempt at a quantitative model of electroweak
baryogenesis, some of which are still awaiting definitive answers.
Nevertheless, it is well established that the transition must be
rather strongly first-order to avoid washing out any previously
generated asymmetry\footnote{We are not considering here the
alternative possibility of a net $(B-L)$ generated before the
electroweak phase transition.}. As an approximate bound for the
single-Higgs case one can take [\ref{shapo}]
\be
\label{nc}
{v(T_c) \over T_c } \simgt 1 \, ,
\ee
where $T_c$ is the temperature at which the symmetric minimum is
degenerate with the symmetry-breaking one, characterized by a vacuum
expectation value $v(T_c)$. In the Standard Model (SM), this
condition turns out to be incompatible with the experimental limit
[\ref{lep}] on the Higgs mass,

\be
\label{exp}
m_h > 60 \gev
\;\;\;\;\;
(95\% \; {\rm c.l.}) \, ,
\ee
even after implementing the presently known techniques
[\ref{standmod},\ref{improved}] for handling the infrared problem.
Simple extensions of the Standard Model, however, could still be
acceptable candidates for electroweak baryogenesis
[\ref{extensions}]. Among them, a particularly attractive one is the
Minimal Supersymmetric extension of the Standard Model (MSSM). In the
MSSM (for a recent review see ref.~[\ref{trieste}]) there can be
extra CP-violating phases [\ref{cpsusy}] besides the
Kobayashi-Maskawa one, which could help in generating the observed
baryon asymmetry [\ref{cn}].  In the following, we shall deal with
the effective potential of the MSSM (for simplicity, in the limit
$\ma \to
\infty$ with $\tb$ fixed, corresponding to only one light Higgs with
SM-like
properties), to check whether conditions (\ref{nc}) and (\ref{exp})
can be
simultaneously satisfied for some acceptable values of the
parameters. Only in
that case can there be room for some particular dynamical mechanism
to
work\footnote{One cannot rigorously exclude that large
non-perturbative effects, not accounted for by the existing
calculations, could modify the predicted values of the sphaleron
energy and of $v(T_c)/T_c$: we have nothing new to say in this
respect, and in the following we shall disregard this possibility.}.
Previous
investigations have been carried out in ref.~[\ref{giudice}], using
the $T$-dependent one-loop potential and assuming a universal soft
supersymmetry-breaking mass for all squarks (other, more
model-dependent considerations have been made in ref.~[\ref{lpu}]).
In this paper, we improve over the previous analyses in two respects.
First, we account for the full structure of the top-squark sector,
which, because of the large top Yukawa coupling, is the only possible
source of important deviations from the SM case, and we discuss the
most important phenomenological constraints on the stop parameters.
Second, we resum the leading
plasma corrections to gauge and scalar boson masses, taking into
account the enlarged particle spectrum of the MSSM: we compute
additional contributions to vector bosons self-energies, as well as
entirely novel effects associated with stop squarks self-energies.

The MSSM contains two complex Higgs doublets, $H_1 \equiv \left(
\begin{array}{c}
H_1^0 \\
H_1^-
\end{array}
\right)$
and $H_2 \equiv \left(
\begin{array}{c}
H_2^+ \\
H_2^0
\end{array}
\right)$, and
its tree-level potential reads, in standard notation [\ref{trieste}]
\bea
\label{v0}
V_0 & = &  m_1^2 \left| H_1 \right|^2 + m_2^2 \left| H_2
\right|^2 + m_3^2 \left( H_1 H_2 + {\rm h.c.} \right) \nonumber \\
& + &  {g^2 \over 8} \left( H_2^{\dagger} {\vec\sigma} H_2
+ H_1^{\dagger} {\vec\sigma} H_1 \right)^2 +
{g'\,^2 \over 8} \left(  \left| H_2 \right|^2 -
\left| H_1 \right|^2 \right)^2 \, .
\eea
It is not restrictive to assume that the only non-vanishing vacuum
expectation values are $v_1 \equiv \langle H_1^0 \rangle$ and $v_2
\equiv \langle H_2^0 \rangle$, both real and positive.  Then, at the
classical level, all Higgs masses and couplings can be expressed in
terms of $\tb \equiv v_2/v_1$ and $\ma = |m_3| \sqrt{\tb + \cot
\beta}$. Zero-temperature quantum corrections can be easily handled
in full generality. At finite temperature, however, the analysis
becomes much more complicated. Besides the obvious difficulty of
studying a $T$-dependent potential in two variables, one might
envisage
the possibility of different non-trivial minima: some of them could
break CP [\ref{comelli}] or other symmetries of the MSSM, and it is
not inconceivable that the phase transition might take place in two
or more steps [\ref{carlson}]. To simplify the problem, we consider
here the limit $\ma \to \infty$, with $\tb$ fixed. In this limit the
low-energy theory contains the single Higgs doublet
\be
\Phi \equiv \cb \; \overline{H}_1 + \sb \; H_2 \, ,
\;\;\;\;\;\;\;\;\;\;
\left( \overline{H}_1 \equiv -i \sigma^2 H_1^* \right) \, ,
\ee
with a tree-level potential of SM form, but with a special value of
the quartic coupling. Calling $\phi/\sqrt{2}$ the constant background
value of the real neutral component of $\Phi$, and restricting our
attention to the $\phi$-dependence, we can write
\be
\label{vzero}
V_0 = - {\mu^2 \over 2} \phi^2 + {\lambda \over 4} \phi^4 \, ,
\ee
with
\be
\lambda = {g^2 + g'\,^2 \over 8} \cos^2 2 \beta \, .
\ee
In this case, the physical Higgs boson has couplings to vector bosons
and fermions of SM strength, and in first approximation the limit of
eq.~(\ref{exp}) still applies.
In general, $m_3^2$ is an independent, soft supersymmetry-breaking
parameter, so that one can formally take the limit $m_A \to \infty$
while keeping all the supersymmetric particle masses finite. To
decide whether this approximation is physically justified, one would
need to know about spontaneous supersymmetry breaking in the
underlying fundamental theory.

\vspace{7 mm}

{\bf 2.}
It is already clear from previous studies [\ref{giudice}] that the
only numerically relevant contributions to the one-loop
finite-temperature effective potential of the MSSM (at the level of
precision required by the problem under consideration) are, besides
the SM ones, those associated with the stop squarks. It is then
important to identify the region of the top-stop parameter space that
is already ruled out by experiment. In the conservative limit of
negligible mixing (as we shall see later, this is the most favourable
situation for baryogenesis), the field-dependent stop and sbottom
masses are given by
\be
\label{smasses}
\begin{array}{ll}
m_{\tilde{t}_L}^2 (\phi) = m_t^2 (\phi) + m_{Q_3}^2 +
D_{\tilde{t}_L}^2 (\phi)  \, ,
&
m_{\tilde{t}_R}^2 (\phi) = m_t^2 (\phi) + m_{U_3}^2 +
D_{\tilde{t}_R}^2 (\phi)  \, ,
\\
&
\\
m_{\tilde{b}_L}^2 (\phi) = m_b^2 (\phi) + m_{Q_3}^2 +
D_{\tilde{b}_L}^2 (\phi)  \, ,
&
m_{\tilde{b}_R}^2 (\phi) = m_b^2 (\phi) + m_{D_3}^2 +
D_{\tilde{b}_R}^2 (\phi)  \, ,
\end{array}
\ee
where $m_{Q_3}$, $m_{U_3}$ and $m_{D_3}$ are soft
supersymmetry-breaking parameters for $(\tilde{t}_L,\tilde{b}_L)$,
$\tilde{t}_R$ and $\tilde{b}_R$, respectively, and ($q=t,b$)
\be
\label{dterms}
D_{\tilde{q}_L}^2(\phi) = m_Z^2 (\phi) \cbb \left[ T_{3L} ( \tilde{q}
) - Q (
\tilde{q} )
\sin^2 \theta_W \right] \, ,
\;\;\;\;\;\;
D_{\tilde{q}_R}^2(\phi) = m_Z^2 (\phi) \cbb \left[Q ( \tilde{q} )
\sin^2 \theta_W \right] \, .
\ee
The field-dependent masses for the gauge bosons and the
third-generation quarks are given by
\be
\label{gauge}
m_W^2 (\phi) = {g^2 \over 4} \phi^2 \, ,
\;\;\;\;\;\;
m_Z^2 (\phi) = {g^2 + g'\,^2 \over 4} \phi^2 \, ,
\ee
and
\be
m_t^2 (\phi) = {h_t^2 \sin^2 \beta \over 2} \phi^2 \, ,
\;\;\;\;\;\;
m_b^2 (\phi) = {h_b^2 \cos^2 \beta \over 2} \phi^2 \, ,
\ee
where $h_t$ and $h_b$ are the top and bottom Yukawa couplings to
$H_2^0$ and $H_1^0$, respectively.

To discuss the experimental constraints, it is appropriate to treat
$m_{Q_3}$, $m_{U_3}$ and $m_{D_3}$ as independent parameters,
unrelated to the
soft supersymmetry-breaking masses of the first two generations.
Specific models for the spontaneous breaking of local supersymmetry
can predict some correlations among $m_{Q_3}$, $m_{U_3}$, $m_{D_3}$
and the other parameters of the MSSM, but we would like here to be as
general as possible. Due to the smallness of $m_b$, bottom and
sbottom loops do not play any important role in the description of
the electroweak phase transition, so that we only need to consider
the constraints on $m_{Q_3}$ and $m_{U_3}$. Nevertheless,
$SU(2)_L$-invariance implies the equality of the soft masses for
$\tilde{t}_L$ and $\tilde{b}_L$: a strong constraint
on their mass then comes from
the fact that the decay $Z \to \tilde{b}_L \tilde{\ov{b}}_L$ has not
been observed at LEP, either directly or via its contribution to the
$Z$-boson width. Since the $Z \tilde{b}_L \tilde{\ov{b}}_L$ coupling
is not particularly suppressed, being proportional to $|-(1/2)+(1/3)
\sin^2 \theta_W| \simeq 0.42$, we can infer
$m_{\tilde{b}_L} \simgt 45 \gev$. Such a result translates into an
excluded region in the $(m_{Q_3},\tb)$-plane, the one to the left of
the
solid line in fig.~1a: contours corresponding to higher values of
$m_{\tilde{b}_L}$ are denoted by dashed lines. Direct squark searches
at hadron colliders [\ref{cdf}] do not provide
model-independent limits on the squark masses of the third
generation. First of all, the published limits assume five or six
degenerate squark flavours. Secondly, even under such an assumption,
the
limit on the squark mass evaporates for a sufficiently large mass of
the gluino
or of the lightest supersymmetric particle. Other, indirect limits on
$m_{Q_3}$
come from electroweak precision measurements, since the
$\tilde{t}_L$--$\tilde{b}_L$ mass
splitting can contribute significantly to the $\rho$ parameter
[\ref{rho}],
\be
\label{deltarho}
\Delta \rho (\tilde{t},\tilde{b}) = {3 g^2 \over 64 \pi^2 m_W^2}
\left( \mstl^2 + \msbl^2 - 2 {\mstl^2 \msbl^2 \over \mstl^2 -
\msbl^2} \log {\mstl^2 \over \msbl^2} \right) \, .
\ee
In the SM, the leading contribution to $\Delta \rho$ comes from the
top-bottom mass splitting,
\be
\label{topbot}
\Delta \rho (t,b) = {3 g^2 \over 64 \pi^2 m_W^2}
\left( m_t^2 +  m_b^2 - 2 {m_t^2 m_b^2 \over m_t^2 -
m_b^2} \log {m_t^2 \over m_b^2} \right) \, ,
\ee
but there are other contributions proportional to $\log (m_h/m_Z)$.
In
the case of a Higgs boson relatively close in mass to the $Z$ boson,
as the one we are considering here, the latter contributions do not
play an important role, and the bound from precision electroweak data
[\ref{fits}] can be taken to be$\Delta \rho (t,b)< 0.008$,
corresponding to
$m_t < 160 \gev$. In the MSSM, there are two types of contributions
to $\Delta
\rho$ (defined in terms of LEP observables) that can give measurable
effects\footnote{Indeed, a refined analysis of the indirect effects
of
relatively light squarks should involve a full one-loop calculation
of the
precisely measured electroweak observables. However, this is beyond
the aim of
the present paper.}. The first type is associated with the vector
boson
self-energies at zero momentum, and includes the terms in
eqs.~(\ref{deltarho})
and (\ref{topbot}). The second type is associated with the $Z$-boson
self-energy at $Q^2 \simeq m_Z^2$: in the presence of supersymmetric
particles
(e.g. charginos) slightly above threshold, there can be large effects
[\ref{barbieri}] that compensate the zero-momentum contribution. Such
cancellation, however, is only effective when considering LEP
observables, but
does not play a role in the fits to low-energy data. To take this
possibility
into account, we shall impose in the following the conservative bound
$\deltarho < 0.01$. When the stop-sbottom
contribution is negligible, this corresponds to $m_t < 180 \gev$.
In the general case, this bound is represented by the solid lines in
fig.~1b: for each indicated value of the top-quark mass, the region
to the left of the corresponding line is excluded. For comparison, we
also draw in fig.~1b dashed lines that correspond to the value
$\deltarho = 0.008$. We are not aware of any model-independent bound
on $m_{U_3}$.

\vspace{7 mm}

{\bf 3.}
We now move to the construction of the one-loop finite-temperature
effective
potential of the MSSM in the chosen limit. Keeping only the
contributions associated with the gauge bosons $W$ and $Z$, with the
quark $t$, and with the squark mass eigenstates
$\st_1$ and $\st_2$, and working in
the 't~Hooft-Landau gauge and in the $\ov{DR}$-scheme, we can write
\be
\label{vnaive}
V_1 = V_0 + \Delta V^{(0)} + \Delta V^{(T)} \, ,
\ee
where $V_0$ has the form of eq.~(\ref{vzero}),
\be
\label{deltav}
\Delta V^{(0)} = {\displaystyle \sum_{i=W,Z,t,\st_1,\st_2} {n_i \over
64 \pi^2} m_i^4 (\phi) \left[ \log {m_i^2 (\phi) \over Q^2} - {3
\over 2} \right] } \, ,
\ee
and
\be
\label{deltavt}
\Delta V^{(T)} = {T^4 \over 2 \pi^2} \left[ \sum_{i=W,Z,\st_1,\st_2}
n_i \, J_+ (y_i^2) + n_t \, J_- (y_t^2) \right] \, .
\ee
In eqs.~(\ref{deltav}) and (\ref{deltavt}),
\be
\label{multi}
n_W=6 \,  , \;\;\;
n_Z=3 \,  , \;\;\;
n_t = - 12 \, ,  \;\;\;
n_{\st_1} = n_{\st_2} = 6 \, ,
\ee
\be
\label{ypsilon}
y^2 \equiv { m^2 (\phi) \over T^2 } \, ,
\;\;\;\;\;
J_{\pm} (y^2) \equiv \int_0^{\infty} dx \, x^2 \,
\log \left( 1 \mp e^{- \sqrt{x^2 + y^2}} \right) \, ,
\ee
and $Q$ is the renormalization scale (we choose for definiteness
$Q^2=m_Z^2$).
For arbitrary mixing in the stop mass matrix\footnote{The
phenomenological constraints on the top-bottom-stop-sbottom sector
discussed before can be trivially generalized to the case of
non-negligible mixing in the stop and sbottom mass matrices.}, the
two field-dependent stop masses are given by
\be
\label{mstop}
m_{\tilde{t}_{1,2}}^2 (\phi) = {m^2_{\tilde{t}_L} (\phi) +
m^2_{\tilde{t}_R} (\phi) \over 2} \mp \sqrt{ \left[
{m^2_{\tilde{t}_L} (\phi)- m^2_{\tilde{t}_R} (\phi) \over 2}
\right]^2 + {m_t^2(\phi) M_t^2 \over \sin^2 \beta} } \, ,
\ee
where $m^2_{\tilde{t}_L} (\phi)$ and $m^2_{\tilde{t}_R} (\phi)$ were
given in eq.~(\ref{smasses}) and $M_t \equiv A_t \sb  + \mu \cb$ is a
mass parameter controlling the mixing in the stop mass
matrix\footnote{In the general case, $M_t=0$ would imply
$A_t=\mu=0$. In the case we are studying, however, $\tan \beta$ is
not a
dynamical variable: $M_t=0$ just defines the line $\mu=-A_t\tan\beta
$ in the
$(\mu,A_t)$ plane, and is consistent with $A_t,\mu \neq 0$.}.

In the chosen limit, the tree-level value of the Higgs mass is
completely determined by $\tb$, $m_h = m_Z |\cos 2 \beta|$, and
the stop masses can be obtained from eq.~(\ref{mstop}) by replacing
$\phi$ with its vacuum expectation value, $v \equiv \langle \phi
\rangle$. To compute the radiatively corrected Higgs mass, and relate
the renormalized parameter $\mu^2$ to physical quantities, one can
easily adapt the calculations of refs.~[\ref{pioneer}]. Defining
$f(m^2) \equiv m^4 [ \log (m^2/Q^2) - (3/2)]$,
one finds
\be
\label{musq}
{\displaystyle
\mu^2 = {m_Z^2 \cos^2 2 \beta \over 2}
+
\sum_i {n_i \over 64 \pi^2}  {1 \over v} {\partial m_i^2(v) \over
\partial v}
f' \left[ m_i^2(v) \right] } \, ,
\ee
and
\be
\label{mhsq}
m_h^2 = m_Z^2 \cos^2 2 \beta + \sum_i {n_i \over 64 \pi^2}  \left\{
\left[ {\partial m_i^2(v) \over \partial v} \right]^2 f'' \left[
m_i^2(v) \right] + \left[ {\partial^2 m_i^2(v) \over \partial v^2} -
{1 \over v} {\partial m_i^2(v) \over \partial v} \right] f' \left[
m_i^2(v) \right] \right\} \, .
\ee

We are now  ready to study the phase transition in the one-loop
approximation. For simplicity, we begin by choosing a universal soft
mass, $m_{Q_3} = m_{U_3} \equiv \tilde{m}$, and negligible mixing,
$M_t=0$, in the stop mass matrix. To illustrate the effects of the
radiative corrections to the Higgs mass, we plot, as dashed lines in
fig.~2a, contours of constant $m_h$ (in GeV) in the $\msqtb$ plane,
for $m_t = 150 \gev$. The pattern of variation of the stop masses is
not particularly interesting: within the intervals of $\msq$ and
$\tb$ considered in fig.~2a, one finds $140 \gev \simlt
m_{\tilde{t}_1} \simlt 250 \gev$, $145 \gev \simlt m_{\tilde{t}_2}
\simlt 250 \gev$. As for the phase transition, we display as solid
lines in fig.~2a contours of constant $v(T_c)/T_c$, obtained from the
potential of eq.~(\ref{vnaive}). We can see that the constraints
(\ref{nc}) and
(\ref{exp}) are satisfied in this case for $\msq \simlt 120 \gev$ and
$\tb \simgt 2.1$. On the other hand, after imposing the further
constraints of fig.~1 we can see that the above bounds do not change
appreciably, but the allowed region gets somewhat reduced. Imposing
the more stringent constraint $\deltarho < 0.008$ would leave no
acceptable region. At this level of approximation, the only
difference with the SM case is the inclusion of the stop
contributions to the one-loop effective potential. With those
contributions removed, one would find for example that $m_h > 60
\gev$ implies $v(T_c)/T_c \simlt 0.6$.

\vspace{7 mm}

{\bf 4.}
It is well known from SM studies that, to obtain a better estimate of
$v(T_c)/T_c$ for the presently allowed values of the Higgs mass, it
is of crucial importance to resum at least the leading
infrared-dominated higher-loop contributions to the $T$-dependent
effective potential, associated with the so-called daisy diagrams
[\ref{standmod},\ref{improved}]. In practice, this amounts to
computing some $T$-dependent effective masses, $\ov{m}_i^2 (\phi,T)$,
for the light bosons of the model under consideration, and to correct
the potential of eq.~(\ref{vnaive}) with the addition of
\be
\label{dvdaisy}
\Delta V_{daisy} = - {T \over 12 \pi} \sum_i n_i \left[ \ov{m}_i^3 (
\phi , T ) - m_i^3 (\phi) \right] \, .
\ee
The sum in eq.~(\ref{dvdaisy}) should run over all the bosons of the
model whose field-dependent masses in the relevant range of $\phi$
values are
not significantly greater than $T_c$. Moreover, one must consider
separately the transverse and longitudinal degrees of freedom of the
gauge bosons, since their propagators are subject to different
$T$-dependent corrections. In the SM, the most important effect is
the Debye screening of the longitudinal gauge bosons.
The inclusion of this screening [\ref{improved}] further reduces the
value of
$v(T_c)/T_c$. For example, in the SM with $m_t = 150 \gev$ and $m_h >
60 \gev$, the daisy-improved effective potential would give
$v(T_c)/T_c \simlt 0.45$.

Similar considerations can be made for the MSSM, but with some
important differences. First of all, loops of supersymmetric
particles can give additional contributions to the effective gauge
boson  masses. Secondly, when $m_{Q_3}, m_{U_3} \ll T$ the
perturbative
expansion
in the stop squark sector, which is controlled by the parameters
\begin{equation}
\label{alpha}
\alpha_{\tilde{t}_1,\tilde{t}_2}^a \equiv
{g^2_a \over 2 \pi} \frac{T^2}{m^2_{\tilde{t}_1,\tilde{t}_2}} \ ,
\;\;\;\;\;\;\;\;\;\;
(g_a=g,g',g_s,h_t)
\ ,
\end{equation}
breaks down at $\phi$ values close to $0$. Including the daisy
diagrams
amounts to a resummation to all orders in
$\alpha_{\tilde{t}_1,\tilde{t}_2}^a$,
and the Debye screening restores the validity of the stop
perturbative
expansion. When $m_{Q_3}, m_{U_3} \gg T$ the perturbative expansion
in the stop
sector is valid for all values of $\phi$. In that case the daisy
resummation is
unnecessary but harmless, since the relative contribution to
(\ref{deltavt})
and (\ref{dvdaisy}) from $\tilde{t}_1$ and $\tilde{t}_2$ drops to
zero. In the
region $m_{Q_3}, m_{U_3} \sim T$, the screening from  $m_{Q_3},
m_{U_3}$
competes with the Debye screening, which has to be taken into
account.
We then conclude that improving the theory by the daisy resummation
is a
consistent procedure\footnote{In this paper we do not consider
subleading
corrections to the daisy approximation, which are controlled by the
parameters
$\beta_{\tilde{t}_1,\tilde{t}_2}^a= (g^2_a/2 \pi) (T/ m_{\tilde{t}_1,
\tilde{t}_2})$.} for the whole range of values of  $m_{Q_3},
m_{U_3}$, and one
has to compute the $T$-dependent self-energies for the stop sector,
which
involve gluon loops as well as loops of supersymmetric particles.

The masses $\ov{m}^2_i$ in (\ref{dvdaisy}) are given by
\be
\label{wmass}
\ov{m}_{W_T}^2  = m_W^2 (\phi) \, ,\hspace{1cm}
\ov{m}_{W_L}^2  = m_W^2 (\phi)+ \Pi_{W_L}  \, ,
\ee
\be
\label{zphtmass}
\ov{m}_{Z_T}^2  = m_Z^2 (\phi) \, , \hspace{1cm}
\ov{m}_{\gamma_T}^2 = 0 \, ,
\ee
\be
\label{zphlmass}
\left(
\begin{array}{cc}
\ov{m}^2_{Z_L} & 0 \\
0 & \ov{m}^2_{\gamma_L}
\end{array}
\right)=R_V
\left(
\begin{array}{cc}
\frac{1}{4} g^2  \phi^2 + \Pi_{W_L} & -\frac{1}{4} g g'  \phi^2  \\
 -\frac{1}{4} g g'  \phi^2 & \frac{1}{4} g' \, ^2  \phi^2 + \Pi_{B_L}
\end{array}
\right)
R^{-1}_V \, ,
\ee
\be
\label{hgmass}
\ov{m}_h^2 = ( 3 \lambda \phi^2 - \mu^2 ) + \Pi_{h} \, , \hspace{1cm}
\ov{m}_{\chi}^2 = ( \lambda \phi^2 - \mu^2 ) + \Pi_{\chi} \, ,
\ee
\be
\label{stopmass}
\left(
\begin{array}{cc}
\ov{m}^2_{\tilde{t}_1} & 0\\
0& \ov{m}^2_{\tilde{t}_2}
\end{array}
\right)=
R_{\tilde{t}}
\left(
\begin{array}{cc}
\ov{m}^2_{\tilde{t}_L} (\phi) + \Pi_{\tilde{t}_L}
& h_t M_t \phi \\
h_t M_t \phi &
\ov{m}^2_{\tilde{t}_R} (\phi) + \Pi_{\tilde{t}_R}
\end{array}
\right) R_{\tilde{t}}^{-1} \, ,
\ee
where we have also included the Higgs ($n_h=1$) and Goldstone
($n_{\chi}=3$) degrees of freedom, and the rotations $R_V$ and
$R_{\tilde{t}}$, which diagonalize the squared mass matrices for the
neutral vector bosons and the stop squarks, respectively. The symbols
$\Pi_i$ denote the (leading) parts of the $T$-dependent self-energies
for the $i^{\rm th}$ boson proportional to $T^2$. They are given by
\be
\label{piw}
\Pi_{W_L} = \frac{9}{4} g^2 T^2 +
\frac{1}{12} g^2 (\cos^4\beta + \sin^4\beta) T^2 +
\left\{ 2 g^2 T^2 \right\} \, ,
\ee
\be
\label{pib}
\Pi_{B_L} = \frac{85}{36} g'\,^2 T^2 +
\frac{1}{12} g'\,^2 (\cos^4\beta + \sin^4\beta) T^2 +
\left\{ \frac{26}{9} g'\,^2 T^2 \right\} \, ,
\ee
$$
\Pi_h = \Pi_{\chi} =
\frac{1}{24} \left[ \frac{3}{4} (g^2+g'\,^2) (\cos^4\beta +
\sin^4\beta)
- 2 g'\,^2 \sin^2\beta\cos^2\beta \right] T^2
$$
\be
\label{pih}
+ \frac{3}{16} g^2 T^2 + \frac{1}{16} g'\,^2 T^2 + \frac{3}{4} h_t^2
\sin^2\beta  T^2+
\left\{\frac{1}{8} g^2 T^2 + \frac{1}{24} g'\,^2 T^2 \right\} \, ,
\ee
\[
\Pi_{\tilde{t}_L} = \frac{4}{9} g_s^2 T^2 + \frac{1}{4} g^2 T^2
+ \frac{1}{108} g'\,^2 T^2 +
\frac{1}{12} h_t^2 (2 + \sin^2\beta) T^2
\]
\be
+\left\{\frac{2}{9} g_s^2 T^2 + \frac{1}{8} g^2 T^2
+ \frac{1}{216} g'\,^2 T^2 \right\} \, ,
\ee
\be
\label{pistr}
\Pi_{\tilde{t}_R} = \frac{4}{9} g_s^2 T^2 + \frac{4}{27} g'\,^2 T^2 +
\frac{1}{6} h_t^2 (2 + \sin^2\beta) T^2 +
\left\{\frac{2}{9} g_s^2 T^2 + \frac{2}{27} g'\,^2 T^2 \right\} \, .
\ee
The previous expressions correspond to {\em maximal} screening, i.e.
to the case where all supersymmetric particles (except the Higgs
bosons
which become superheavy in the limit $m_A \rightarrow \infty$)
have masses smaller
than or of the order of the critical temperature, and thus contribute
to the self-energies. Terms outside curly brackets correspond to the
contribution from the SM particles and the squarks of the third
generation.
Terms in curly brackets correspond to the contribution from
supersymmetric
particles other than the squarks of the third generation, i.e.
sleptons, squarks of the first and second generations, charginos,
neutralinos and gluinos. The self-energies for the case of {\em
minimal}
screening, i.e. when the only light supersymmetric particles are the
squarks of the third generation, can be
read off (\ref{piw})--(\ref{pistr}) by removing the terms in curly
brackets. Since gluinos provide the main contribution, $O(g_s^2)$,
from
the terms inside curly brackets, the cases of maximal and minimal
screening
can be thought of as cases of {\em light} and {\em heavy} gluinos.
In particular, in the case of minimal screening the inclusion of
light neutralinos, charginos, sleptons and squarks would not alter
significantly the corresponding numerical results, and scenarios
where the lightest supersymmetric particle is a light neutralino
or a sneutrino are not excluded by our analysis.

The effects of including the resummation of the daisy diagrams, for
the cases of maximal and minimal screening, are illustrated in
figs.~2b and 2c, respectively. One can see that the allowed region
can shrink considerably with respect to the one-loop approximation.
In the case of maximal screening (which we use to emphasize the
effect, but is not likely to correspond to a real situation) no
acceptable points are left in the $\msqtb$-plane. In the case of
minimal screening, which should be interpreted as our best guess for
a realistic situation, the acceptable region is reduced to a small
triangle corresponding to $20 \gev \simlt \tilde{m} \simlt 85 \gev$
and $2 \simlt \tb \simlt 2.8$. Again, no acceptable region would be
left if we had assumed the more stringent bound $\deltarho < 0.008$.

We now want to discuss the dependence of the results on the top-quark
mass and on the stop mixing parameter, which is illustrated in
figs.~3 and 4, respectively. In figs.~3a and 4a, solid lines
correspond to the stop squark masses, the dashed lines to the Higgs
mass, and the remaining free parameters are fixed to the
representative values $\msq = 50 \gev$, $\tb = 2.25$, $M_t=0$
(fig.~3a) and $m_t = 150 \gev$ (fig.~4a). The dependences of
$v(T_c)/T_c$ are illustrated in figs.~3b and 4b
(solid lines), respectively: one
can see that the most
favourable situations are obtained for large top mass and small stop
mixing. As for the effect of the top-quark mass, however, one should
keep in mind the competing effect coming from the bound on $\Delta
\rho$, which disfavours situations with large $m_t$ and small
$\tilde{m}$. Using as before the conservative bound $\Delta \rho <
0.01$, and working in the approximation of minimal screening, we
obtain solutions for $130 \gev \simlt m_t \simlt 160 \gev$. Had we
used the more stringent bound $\deltarho < 0.008$, we would have
found solutions only for $130 \gev \simlt m_t \simlt 140 \gev$.

Before concluding we would like to discuss the uncertainty in the
definition of the critical temperature $T_c$ and its influence on
the calculation of $v(T_c)/T_c$. We have defined $T_c$
as the temperature at which the symmetric minimum is degenerate with
the symmetry-breaking one. This definition does correspond to
the onset of the phase transition, though the latter will take place
at a
somewhat lower temperature
$T_{{\rm ph}}$. On the other hand, the end of the phase transition
happens
at a second critical temperature $T'_c$, which is defined as the
temperature at which the curvature of the effective potential
vanishes at the origin. It is also clear that
$T'_c \simlt T_{{\rm ph}}$, and so
\begin{equation}
\label{diff}
\frac{v(T_c)}{T_c} \simlt \frac{v(T_{{\rm ph}})}{T_{{\rm ph}}}
\simlt  \frac{v(T'_c)}{T'_c}  \, .
\end{equation}
We plot the values of $v(T'_c)/T'_c$ as dashed lines in figs.~3b and
4b. The
quantity $v(T_{{\rm ph}})/T_{{\rm ph}}$ lies in the stripe between
the solid
and dashed lines. We have adopted in this paper the
bound of eq.~(\ref{nc}), which is consistent with the bound
$v(T'_c)/T'_c
\simgt 1.3$, which was used in some of the previous analyses
[\ref{shapo},\ref{improved},\ref{extensions},\ref{giudice}].

Finally, we need to discuss the effects of relaxing the assumption of
a universal soft mass in the stop sector. In this case one can take
advantage of the fact that $m_{U_3}$ can be pushed to very small
values without violating any bound. The most favourable situation is
the one with large $m_{Q_3}$, so that the contribution to $\Delta
\rho$ is negligible, and very small $m_{U_3}$. One can then find
solutions up to the value of $m_t$ that saturates the bound on
$\Delta \rho$.

\vspace{7 mm}

{\bf 5.}
To summarize our results, the region of the MSSM parameter space that
is still compatible with the scenario of electroweak baryogenesis and
with the present experimental constraints is quite restricted. It
seems unlikely that it can be matched with universal boundary
conditions on soft supersymmetry-breaking masses at the
grand-unification scale, as is often assumed by model builders.
A thorough and reliable discussion of this point, however, would be
rather complicated. Moreover, there is no strong theoretical argument
to forbid deviations from universality, especially in the
stop-sbottom sector. In the case of general supersymmetry breaking
considered in this paper, the allowed
region corresponds to a rather light spectrum in the Higgs and
stop-sbottom sectors, $m_h \simlt 70 \gev$ and $m_{\tilde{b}_1}
\simlt 105 \gev$: these predictions should be definitely tested at
LEP~II in the forthcoming years.
Our results are quite sensitive to the assumed bound on $\Delta
\rho$: as is clear from fig.~1b, more stringent bounds on $\Delta
\rho$ drastically reduce the allowed region of parameter space.
Finally, we would like to recall that our analysis was carried out in
the limit of a single light Higgs: as already mentioned, the case of
small $m_A$ is much more cumbersome to analyse and goes beyond the
aim of the present paper.

\section*{Acknowledgements}

We would like to thank R.~Barbieri, F.~Caravaglios and
M.E.~Shaposhnikov for
useful discussions.

\newpage

\section*{References}
\begin{enumerate}
\item
\label{reviews}
M.E.~Shaposhnikov, in `Proceedings of the 1991 Summer School in High
Energy Physics and Cosmology', Trieste, 17 June--9 August 1991,
E.~Gava et al. eds. (World Scientific, Singapore, 1992), Vol.~1,
p.~338; A.G.~Cohen, D.B.~Kaplan and A.E.~Nelson, San Diego preprint
UCSD-PTH-93-02, BUHEP-93-4.
\item
\label{shapo}
M.E.~Shaposhnikov, JETP Lett. 44 (1986) 364; Nucl. Phys. B287 (1987)
757 and B299 (1988) 797; A.I.~Bochkarev, S.Yu.~Klebnikov and
M.E.~Shaposhnikov, Nucl. Phys. B329 (1990) 490; M.~Dine, P.~Huet and
R.~Singleton Jr., Nucl. Phys. B375 (1992) 625.
\item
\label{lep}
J.-F.~Grivaz, Orsay preprint LAL 92-59, and references therein.
\item
\label{standmod}
S.~Weinberg, Phys. Rev. D9 (1974) 3357; L.~Dolan and R.~Jackiw, Phys.
Rev. D9 (1974) 3320; D.A.~Kirzhnits and A.D.~Linde, Zh. Eksp. Teor.
Fiz.
67 (1974) 1263 and Ann. Phys. 101 (1976) 195; P.~Fendley, Phys. Lett.
B196
(1987) 175.
\item
\label{improved}
M.E.~Carrington, Phys. Rev. D45 (1992) 2933; M.~Dine, R.G.~Leigh,
P.~Huet, A.~Linde and D.~Linde, Phys. Lett. B283 (1992) 319 and Phys.
Rev. D46 (1992) 550; P.~Arnold, Phys. Rev. D46 (1992) 2628;
J.R.~Espinosa, M.~Quir\'os and F.~Zwirner, Phys. Lett. B291 (1992)
115 and preprint CERN-TH.6577/92, IEM-FT-58/92; C.G.~Boyd, D.E.~Brahm
and S.D.~Hsu, Caltech preprint CALT-68-1795, HUTP-92-A027, EFI-92-22;
W.~Buchm\"uller and T.~Helbig, preprint DESY 92-117; W.~Buchm\"uller,
T.~Helbig and D.~Walliser, preprint DESY 92-151;
P.~Arnold and O.~Espinosa, Univ. of Washington preprint UW/PT-92-18;
W.~Buchm\"uller, Z.~Fodor, T.~Helbig and D.~Walliser, preprint DESY
93-021.
\item
\label{extensions}
A.I.~Bochkarev, S.V.~Kuzmin and M.E.~Shaposhnikov, Phys. Lett. B244
(1990) 275 and Phys. Rev. D43 (1991) 369; N.~Turok and J.~Zadrozny,
Phys. Rev. Lett. 65 (1990) 2331, Nucl. Phys. B358 (1991) 471 and B369
(1992) 728; L.~Mc~Lerran, M.~Shaposhnikov, N.~Turok and M.~Voloshin,
Phys. Lett. B256 (1991) 451; B.~Kastening, R.D.~Peccei and X.~Zhang,
Phys. Lett. B266 (1991) 413; G.W.~Anderson and L.J.~Hall, Phys. Rev.
D45 (1992) 2685; M.~Dine, P.~Huet, R.~Singleton Jr. and L.~Susskind,
Phys. Lett. B257 (1991) 351; K.~Enqvist, K.~Kainulainen and I.~Vilja,
Phys. Rev. D45 (1992) 3415; N.~Sei, I.~Umemura and K.~Yamamoto, Phys.
Lett. B299 (1993) 286; M. Pietroni, Padova preprint DFPD/92/TH/36;
J.R.~Espinosa and M.~Quir\'os, Madrid preprint
IEM-FT-67/93, to appear in Phys. Lett.~B.
\item
\label{trieste}
F.~Zwirner, in `Proceedings of the 1991 Summer School in High Energy
Physics and Cosmology', Trieste, 17 June--9 August 1991, E.~Gava et
al. eds. (World
Scientific, Singapore, 1992), Vol.~1, p.~193.
\item
\label{cpsusy}
J.~Ellis, S.~Ferrara and D.V.~Nanopoulos, Phys. Lett. B114 (1982)
231; W.~Buchm\"uller and D.~Wyler, Phys. Lett. B121 (1983) 321;
J.~Polchinski and M.B.~Wise, Phys. Lett. B125 (1983) 393;
F.~del~Aguila, B.~Gavela, J.~Grifols and A.~Mendez, Phys. Lett. B126
(1983) 71; D.V.~Nanopoulos and M.~Srednicki, Phys. Lett. B128 (1983)
61; M.~Dugan, B.~Grinstein and L.J.~Hall, Nucl. Phys. B255 (1985)
413.
\item
\label{cn}
A.G.~Cohen and A.E.~Nelson, Phys. Lett. B297 (1992) 111.
\item
\label{giudice}
G.F.~Giudice, Phys. Rev. D45 (1992) 3177;
S.~Myint, Phys. Lett. B287 (1992) 325.
\item
\label{lpu}
D.V.~Nanopoulos and H.~Pois, preprint CERN-TH.6763/92.
\item
\label{comelli}
D.~Comelli and M.~Pietroni, Padova preprint DFPD-93/TH/06,
UTS-DFT-93-1.
\item
\label{carlson}
D.~Land and E.D.~Carlson, Phys. Lett. B292 (1992) 107;
E.D.~Carlson and W.D.~Garretson, Harvard preprint HUTP-92/A067.
\item
\label {pioneer}
Y.~Okada, M.~Yamaguchi and T.~Yanagida,
Prog. Theor. Phys. Lett. 85 (1991) 1;
J.~Ellis, G.~Ridolfi and F.~Zwirner, Phys. Lett. B257 (1991) 83;
R.~Barbieri and M.~Frigeni, Phys. Lett. B258 (1991) 395;
J.~Ellis, G.~Ridolfi and F.~Zwirner, Phys. Lett. B262 (1991) 477.
\item
\label{cdf}
F.~Abe et al. (CDF Collaboration), Phys. Rev. Lett. 69 (1992) 3439.
\item
\label{rho}
L.~Alvarez-Gaum\'e, J.~Polchinski and M.~Wise, Nucl. Phys. B221
(1983) 495;
R.~Barbieri and L.~Maiani, Nucl. Phys. B224 (1983) 32;
C.S.~Lim, T.~Inami and N.~Sakai, Phys. Rev. D29 (1984) 1488.
\item
\label{fits}
L.~Rolandi, preprint CERN-PPE-92-175, and references therein.
\item
\label{barbieri}
R.~Barbieri, M.~Frigeni and F.~Caravaglios, Phys. Lett. B279 (1992)
169;
J.~Ellis, G.L.~Fogli and E.~Lisi, preprint CERN-TH.6643/92.
\end{enumerate}
\vfill{
\section*{Figure captions}
\begin{itemize}
\item[Fig.1:]
Exclusion contours in the $(m_{Q_3},\tb)$ plane, for negligible stop
mixing,
corresponding to different experimental constraints: a)~the solid
line corresponds to  $\msbl = 45 \gev$, the dashed lines to the
indicated higher values of $\msbl$ (in GeV); b)~for each indicated
value of $m_t$ (in GeV), the solid lines correspond to $\Delta \rho =
0.01$, the dashed lines to $\Delta \rho = 0.008$.

\item[Fig.2:]
Lines of constant $v(T_c)/T_c$ (solid) and $m_h$ (dashed, in GeV) in
the $(m_{Q_3},\tb)$ plane, for $m_{Q_3} = m_{U_3} \equiv \msq$, $m_t
=
150 \gev$ and $M_t=0$: a)~one-loop approximation; b)~daisy-improved,
for maximal screening; c)~daisy-improved, for minimal screening.

\item[Fig.3:]
Dependence on $m_t$ of some relevant quantities in the representative
case $m_{Q_3} = m_{U_3} \equiv \msq = 50 \gev$, $\tb = 2.25$, $M_t =
0$: a) Higgs (dashed line) and stop masses (solid line), in GeV; b)
$v(T_c)/T_c$ (solid line) and $v(T'_c)/T'_c$ (dashed line),
daisy-improved for minimal screening.

\item[Fig.4:]
Dependence on $M_t$ of some relevant quantities in the representative
case $m_t = 150 \gev$, $m_{Q_3} = m_{U_3} \equiv \msq = 50 \gev$,
$\tb = 2.25$: a) Higgs (dashed line) and stop masses (solid line), in
GeV; b) $v(T_c)/T_c$ (solid line) and $v(T'_c)/T'_c$ (dashed line),
daisy-improved for minimal screening.
\end{itemize}
}
\end{document}